\documentclass[showpacs,aps,floats,epsf,psfig,prc,twocolumn,graphics]{revtex4}
\usepackage{graphicx}
\usepackage{amssymb}
\usepackage{epstopdf}
\usepackage{subfigure}
\def\cnb    {C$\nu$B}

\newcommand{\mitt}{Laboratory for Nuclear Science, Massachusetts Institute of Technology, Cambridge, MA}
\newcommand{\ucsb}{Department of Physics, University of California Santa Barbara, Santa Barbara, CA }

\begin{document}

\title{Sensitivity of Neutrino Mass Experiments to the Cosmic Neutrino Background}

\author{A.~Kaboth}\affiliation{\mitt}
\author{J.~A.~Formaggio}\affiliation{\mitt}
\author{B.~Monreal}\affiliation{\ucsb}

\begin{abstract}
The KATRIN neutrino experiment is a next-generation tritium beta decay experiment aimed at measuring the mass of the electron neutrino to better than 200 meV at 90\% C.L.  Due to its intense tritium source, KATRIN can also serve as a possible target for the process of neutrino capture, $\nu_e + ^3$H$\rightarrow~^3$He$^+ + e^-$.  The latter process, possessing no energy threshold, is sensitive to the Cosmic Neutrino Background (\cnb).  In this paper, we explore the potential sensitivity of the KATRIN experiment to the relic neutrino density.  The KATRIN experiment is sensitive to a \cnb~over-density ratio of $2.0 \times 10^{9}$ over standard concordance model predictions (at 90\% C.L.), addressing the validity of certain speculative cosmological models.
\end{abstract}


\date{\today}

\maketitle

\section{Introduction}
\label{sec:Intro}

The observation of the Cosmic Microwave Background (CMB) stands as one of the most significant scientific discoveries of the 20th century~\cite{penzias1965mea,Dicke:1965zz}.  The series of measurements made after its initial observation---ever increasing in precision and scope---has helped transform observational cosmology into an extremely predictive science~\cite{smoot1992scd}.  One need only to look at the recent discoveries of dark matter and dark energy to appreciate this progression\cite{spergel2003fyw,peebles2003cca}.  Many of the parameters of the cosmological concordance model have now been measured and further exploration of these parameters over the next few decades are likely to reveal even more hidden aspects of our universe.

Though many of the predictions from cosmology have now been realized, there are others that remain unobserved.  Among these is the observation of the Cosmic Neutrino Background (\cnb) produced from the primordial Big Bang.  Like their photon counterparts, their existence in the cosmos is expected; yet, direct observation of \cnb~remains elusive.  Observation of the cosmic relic neutrinos ---or, conversely, the absence of such --- stands as an important verification of the theory.  

Direct observation of relic neutrinos is an extremely difficult challenge from an experimental perspective.  The neutrino temperature $T_\nu$ is related directly to the CMB temperature, $T_\gamma$, such that $T_\nu = (4/11)^{\frac{1}{3}}T_\gamma$, or \mbox{$\sim$1.9~K~\cite{Mather:1999zz}}.  The average \cnb~neutrino kinetic energy, assuming a simple Fermi-Dirac distribution, is expected to be less than 0.2 meV.   Most conventional methods of detecting neutrinos, such as water Cerenkov or scintillator detectors, rely on interactions that have some threshold for the energy of the incoming neutrino, which is often many orders of magnitude larger than the \cnb~energies.  Fortunately, there does exist a promising candidate by which such low energy neutrinos can be detected: neutrino capture on radioactive nuclei.  The process was originally proposed by Weinberg~\cite{weinberg1962und} but has been recently revisited by others~\cite{cocco2007ple, lazauskas2008ccc}.  The generic process, $\nu_e + N \rightarrow N' + e^{-}$, is energetically allowed when $m_\nu + m_N > m_{N'}+m_e$. Conveniently, this is also the required energetics at very small neutrino mass for the more familiar process of beta decay. The signal of this neutrino capture process is distinctive:  the electrons create a monoenergetic peak at $E = E_0 + m_\nu$, where $E$ is the kinetic energy of the detected electron, $E_0$ is the endpoint energy of the beta decay process assuming zero neutrino mass and $m_\nu$ is the mass of the electron neutrino.   For a detector with sufficient resolution, it is in principle possible to separate this monoenergetic peak from the tail distribution of the beta decay spectrum.

Recent and upcoming experiments that make use of beta decay nuclei have made significant advances in both source strength and energy resolution in order to  probe the scale of neutrino masses.  It is natural to inquire, therefore, what sensitivity can reasonably be expected on the \cnb~from these experiments given their experimental and physical limitations.  In this paper, we explore the potential sensitivity of tritium beta decay experiments, like the KArlsruhe TRItium Neutrino (KATRIN) experiment, to cosmic relic neutrinos~\cite{angrik2005kdr}. 

\section{Detection of the \cnb}
\label{sec:cnb}

The total \cnb~rate depends on the cross-section of $\nu$-capture on tritium and the local \cnb~density.

\begin{equation}
\label{eq:rate}
R_{\nu} = n_\nu \frac{N_A~M_{\rm eff}}{A} \int \sigma_{\nu } v_\nu f(p_\nu) \frac{d^3p_\nu}{2\pi^3}
\end{equation}

\noindent where $N_A$ is Avogadro's number, $A$ is the target atomic number, $n_\nu$ is the relic neutrino density, $M_{\rm eff}$ is the effective target mass, $\sigma_{\rm}$ is the \cnb~cross-section, $v_\nu$ and $p_\nu$ are the neutrino velocity and momenta, respectively, and $f(p_\nu)$ is the momentum distribution of the relic neutrinos, which we treat as a simple Fermi-Dirac distribution of characteristic temperature $T_\nu$.  The expected cross-section for relic neutrinos on tritium in the limit of zero momentum yields $\sigma_\nu \frac{v}{c}$ of $(7.84 \pm 0.03) \times 10^{-45}$cm$^2$\cite{cocco2007ple}. The relatively high cross-section places tritium as a preferred target for the sought reaction.  The other commonly used beta decay isotope, $^{187}$Re, is employed by the MARE experiment~\cite{Monfardini:2005dk} and is valued for its low endpoint energy.  However, the cross-section for neutrino capture on $^{187}$Re is almost seven orders of magnitude lower than that of tritium.  Other practical considerations, such as availability, the low energy of the endpoint, and final state effects make tritium favorable in comparison to other potential target materials.

The detected neutrino capture reaction also depends on the {\it local} neutrino density.  A local enhancement of the neutrino density --over the standard 55 neutrinos/cm$^3$ per flavor-- can occur if the neutrino mass is large enough to allow clumping within the Milky Way.  The amount of neutrino over-density will in general depend on both the neutrino mass and the matter density profile of our galaxy.  Table~\ref{tab:rates} shows the expected \cnb~capture rate given KATRIN's tritium target mass under a number of density scenarios:  (a) a standard homogenous Fermi-Dirac distribution; (b) a Navarro, Frenk and White~\cite{white1996} dark matter halo profile, and (c) and for a mass distribution of the Milky Way (MW), using the local neutrino densities computed in~\cite{ringwald2004hep}.  Clearly, the expected rates shown here are well below the detection threshold. However, understanding the analysis and limits at the KATRIN experiment provides a critical first step in understanding how to build and develop future \cnb~experiments. 

\begin{table}[htbp]
   \centering
   \begin{tabular}{|c|c|c|c|} 
	\hline
	&\multicolumn{3}{c|}{Event Rates (events/yr)}\\
	\hline
	$m_{\nu}$&Fermi-Dirac&Navarro, Frenk, \& White& Milky Way\\
	\hline
	0.6& 5 $\times 10^{-6}$& 6.0 $\times 10^{-5}$& 1.0 $\times 10^{-4}$\\
	0.3& 5 $\times 10^{-6}$ & 1.5 $\times 10^{-5}$& 2.2 $\times 10^{-5}$\\
	0.15& 5 $\times 10^{-6}$& 6.7 $\times 10^{-6}$& 8.0 $\times 10^{-6}$ \\
	\hline
   \end{tabular}
   \caption{The event rates for three different neutrino masses and three different mass profiles for the \cnb. Rates are calculated by scaling the results of Ref~\cite{cocco2007ple} by the tritium mass of the KATRIN experiment. All rates are given in events/yr. }
   \label{tab:rates}
\end{table}

\section{The KATRIN Neutrino Mass Experiment}
\label{sec:KATRIN}

The KArlsruhe TRItium Neutrino (KATRIN) experiment is the next generation tritium beta decay experiment with sub-eV sensitivity to make a direct, model independent measurement of the electron neutrino mass. The principle of the experiment is to look for a distortion at the high energy endpoint of the electron spectrum of tritium $\beta$-decay:

\begin{equation}
^{3}{\rm H} \rightarrow~ ^{3}{\rm He}^+ + e^- + \bar \nu_e.
\end{equation}

The shape of the electron energy spectrum of tritium beta decay is determined by well-understood or measurable quantities. Any deviation from this shape would be directly attributable to neutrino mass and would allow a direct determination of the mass of the electron neutrino. After three years of running, KATRIN will be able to achieve a sensitivity of $m(\nu) < 200$ meV (at 90\% C.L.).  This level represents an order of magnitude improvement on the absolute neutrino mass scale. Since the measurement of electrons extends beyond the endpoint of the decay, in order to monitor background, it is also possible to look for the \cnb~signature. 

The experiment is located at the Forschungszentrum Karlsruhe (FZK). The FZK is a unique location as it hosts the Tritium Laboratory Karlsruhe, which is in charge of the tritium cycle of the international ITER fusion program. It thus offers the expertise needed to handle large quantities of tritium. KATRIN will use the windowless gaseous tritium source technique, as used by Los Alamos~\cite{robertson1991lea} and Troitsk~\cite{lobashev2001dsn}. Decay electrons from the source pass through a 10-meter long differential and cryogenic pumping subsection guided by superconducting magnets. The purpose of the differential pumping system is to prevent gas from entering the spectrometer system, which would degrade resolution and raise background by contaminating the system with tritium. 

The KATRIN experiment is based on technology developed by the Mainz~\cite{kraus2005frp} and Troitsk~\cite{lobashev2001dsn} tritium beta decay experiments. These experiments used a so-called MAC-E-Filter (Magnetic Adiabatic Collimation combined with an Electrostatic filter). This technology draws the isotropic electrons from a decay or capture event along magnetic field lines through a decreasing magnetic field so that the cyclotron motion of the electrons around the magnetic field lines is transformed into longitudinal motion along the magnetic field lines. A retarding potential is applied such that only electrons with energy greater than the retarding potential are transmitted to an electron counting detector. By varying the retarding potential, the shape of the decay spectrum can be reconstructed. The energy resolution of this measurement is determined by the ratio $\frac{\Delta E}{E} = \frac{B_A}{B_{\rm max}} = \frac{1}{20000}$, where $B_A$ is the magnetic field in the analyzing plane (the point of maximum potential) and $B_{\rm max}$ is the maximum magnetic field. The decay electrons exiting the spectrometer are imaged onto a silicon PIN diode array using a 6 T superconducting magnet. The electrons counted in the detector array comprise the signal needed to reconstruct the beta-decay spectrum.

The most critical aspect of the KATRIN experiment relevant for the detection of the \cnb~is the number of tritium atoms available for neutrino capture and electron detection. This is given by \[ N_{eff} = N(T_2) \cdot \frac{\Omega}{2\pi} \cdot P \] where $N(T_2)$ is the number of tritium molecules, $\frac{\Omega}{2\pi}$ is the solid angle of the source as seen by the detector, and $P$ is the probability that the emitted electron exits of the source without undergoing an inelastic scattering process.

The number of tritium molecules in the source is given by \[ N(T_2) = \rho d \cdot \epsilon_T \cdot A_s \]  where $\rho d$ is the source column density, $\epsilon_T=0.95$ is the tritium purity, and $A_s=52.65$ cm$^2$ is the source area. Thus the signal rate can be written as \[ N_{eff} = A_s \cdot \epsilon_T \cdot \frac{\Omega}{2\pi} \cdot P \cdot \rho d \] The factor $P \cdot \rho d$ can be considered as an effective column density, $\rho d_{eff}$, which has the value $3.58\times10^{17}/$cm$^2$ at KATRIN.~\cite{angrik2005kdr}  The effective tritium source strength is equivalent to $N_{eff} = 6.64\times 10^{18}$ tritium molecules, or an equivalent mass of $M_{\rm eff} = 66.5 \mu$g.

Because the KATRIN experiment measures the integral beta decay spectrum above some threshold $qU$, the electron spectrum is really the convolution of the $\beta$ and \cnb~electron spectrum, $dN/dE$, and the transmission function of the detector, $T(E,qU)$.  KATRIN also expects a small but finite background rate, $N_b$, to contribute to the overall signal.  Currently, this background rate is expected to be of order 10 mHz in the signal region of interest, independent of retarding voltage.

\begin{equation}
G(qU) = \int_{qU}^{\infty} \frac{dN}{dE} T(E,qU) dE + N_b
\end{equation}

The tritium beta decay energy spectrum has an analytic form~\cite{Lobashev1985305} given by Eq.~\ref{eq:betadecay}

\begin{widetext}
\begin{equation}
	\frac{dN}{dE} =  \sum_{fs} (N_{T_2}F(Z,E)p_e(E+m_e)(E_0-E)\sqrt{(E_0-E)^2 -m_{\nu}^2}\Theta(E_0-E(fs)-m_{\nu}) + N_{C\nu B} e^{\frac{-(E_0-E(fs)+m_{\nu})^2}{2\sigma^2}})P(fs)
	\label{eq:betadecay}
\end{equation}
\end{widetext}

\noindent where $F(Z,E)$ is the Fermi function for beta decay, $E_0$ is the endpoint energy of the $^3$H$_2 \rightarrow (^3$He$^3$H)$^+ + e^- +\bar{\nu}_e$ decay, $E$ is the kinetic energy of the emitted electron, and $N_{T_2}$ and $N_{C\nu B}$ are the rates of tritium beta decay and neutrino capture, respectively. The gaussian term represents the capture signal from the \cnb~centered at one neutrino mass above the endpoint, with a width $\sigma$ chosen to be smaller than any characteristic resolution present in the experiment, but sufficiently large to be reliably integrable by numerical methods.  Since the target involves the presence of molecular T$_2$ gas, one must include any corrections to the endpoint energy due to the molecular daughter molecule following the tritium decay. An accounting of these states is given in~\cite{saenz2000imf}.  Of most relevance are the effects of the rotational-vibrational contributions from decays to the ground state, which introduce a mean excitation energy of 1.7 eV with an inherent broadening of 0.36 eV.  In this analysis, the final states are taken into account via a summation over the states $fs$ of the He$^+$T molecule, each final state weighted by the probability $P(fs)$ for that state occurring.

\begin{figure}[htbp] 
\centering
\includegraphics[width=0.95\columnwidth,keepaspectratio=true]{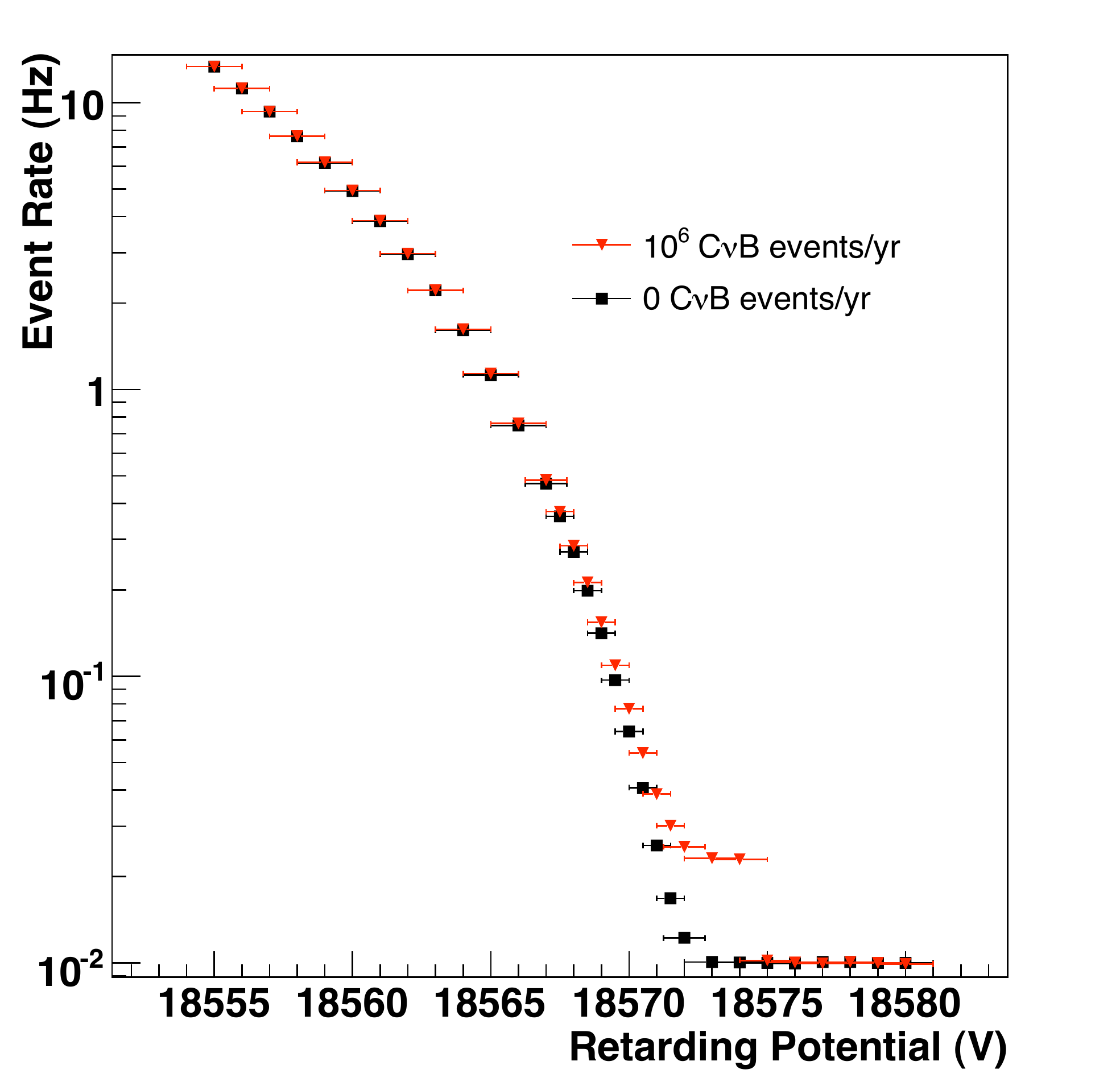} 
\caption{The anticipated beta decay spectrum as a function of retarding voltage with (black) and without (red) neutrino capture events.  Neutrino mass is assumed to be 1 eV.}
\label{fig:spectrum}
\end{figure}

The transmission function, $T(E,qU)$, depends on the value of the retarding potential, $qU$,  as well as the intrinsic resolution of the main spectrometer. For an isotropic source, $T(E,qU)$ can be written analytically as:

\begin{equation}
	T(E,qU)= \left\{ \begin{array}{ll}
         0 & \mbox{if $E-qU <  0$}\\
	\frac{1-\sqrt{1-\frac{(E-qU) B_S}{E B_A}}}{1-\sqrt{1-\frac{\Delta E B_S}{E B_{\rm max}}}} & \mbox{if $0 \leq E-qU \leq \Delta E$}\\
         1 & \mbox{if $E-qU > \Delta E$}.\end{array} \right.
	\label{eq:transmission}
\end{equation}

\noindent where E is the electron energy, $B_S$ is the magnetic field at the source, $B_A$ is the magnetic field at qU, $B_{\rm max}$ is the maximum (pinch) field, and $\frac{\Delta E}{E} = \frac{1}{20000}$ at KATRIN.   A sample decay spectrum, with and without neutrino capture, is shown in Figure~\ref{fig:spectrum}.

\section{Signal Extraction and Results}
\label{sec:signal}

\begin{figure}[htbp] 
   \begin{tabular}{c c}
   \subfigure[$m_{\nu}=0.0$~eV]{\label{fig:mnu0_0}\includegraphics[width=0.5\columnwidth,keepaspectratio=true]{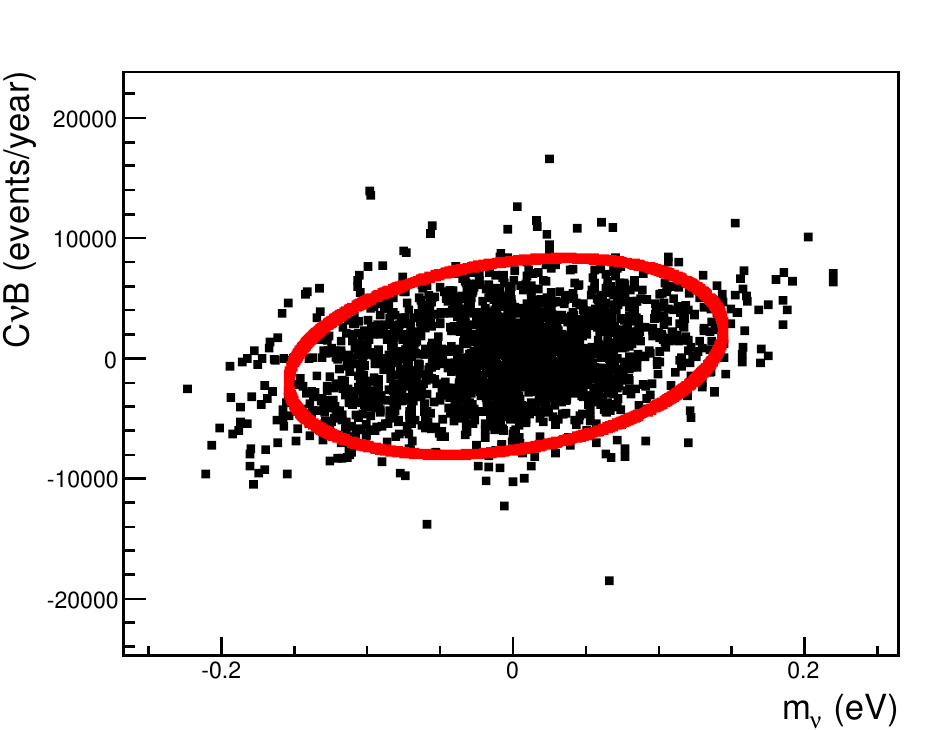}} &
   \subfigure[$m_{\nu}=0.3$~eV]{\label{fig:mnu0_3}\includegraphics[width=0.5\columnwidth,keepaspectratio=true]{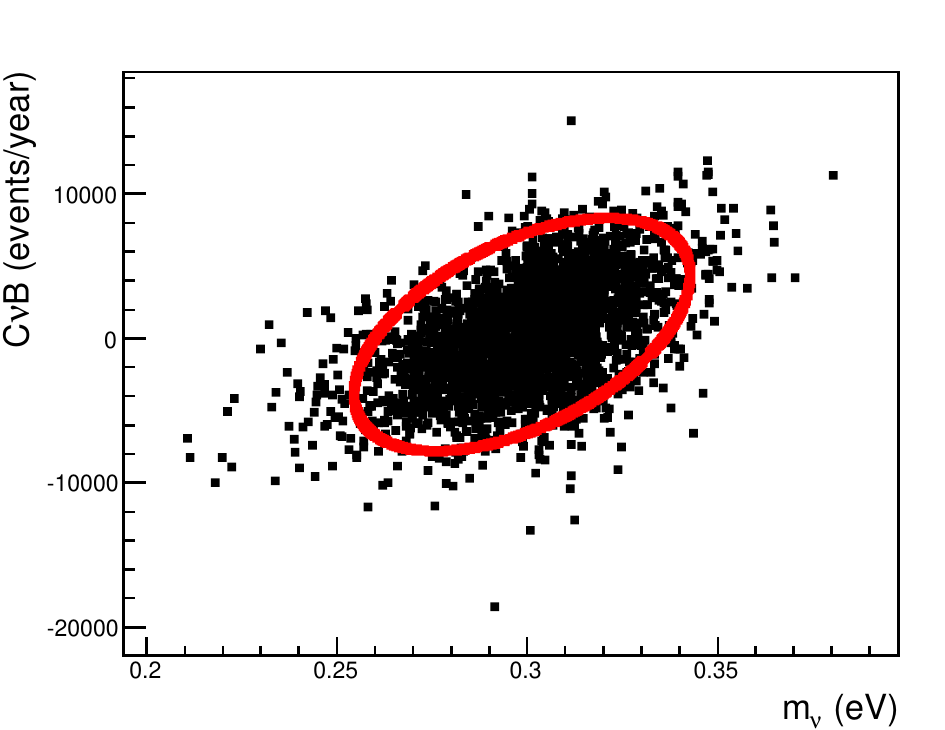}} \\
   \subfigure[$m_{\nu}=0.6$~eV]{\label{fig:mnu0_6}\includegraphics[width=0.5\columnwidth,keepaspectratio=true]{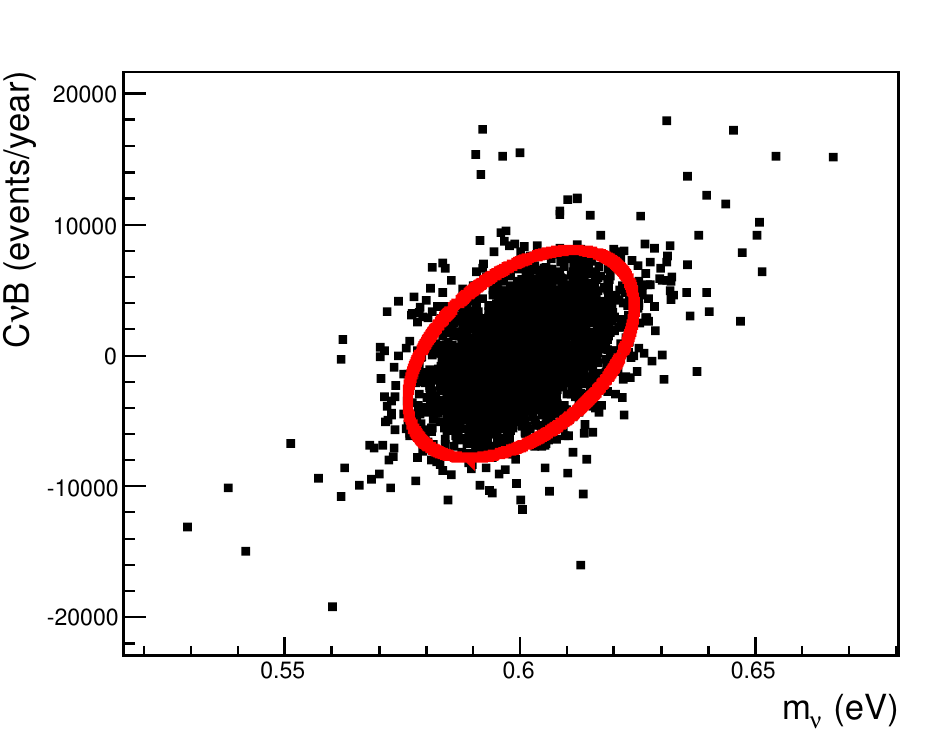}} &  
   \subfigure[$m_{\nu}=1.0$~eV]{\label{fig:mnu1_0}\includegraphics[width=0.5\columnwidth,keepaspectratio=true]{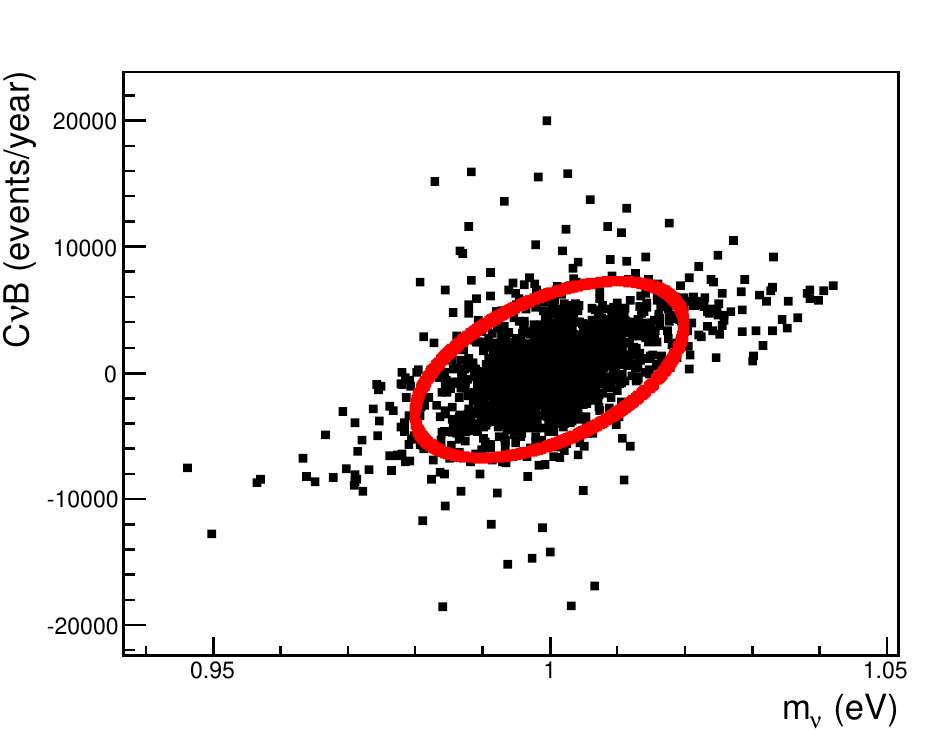}} \\
   \end{tabular}
   \caption{Confidence regions for cosmic neutrino captures in events per year versus neutrino mass in eV for four example neutrino masses. Statistical errors only are shown. Red ellipse shows 90\% C.L in the \cnb~events per year and neutrino mass parameter space.}
   \label{fig:contours}
\end{figure}

\begin{figure}[htbp] 
\centering
\includegraphics[width=0.95\columnwidth,keepaspectratio=true]{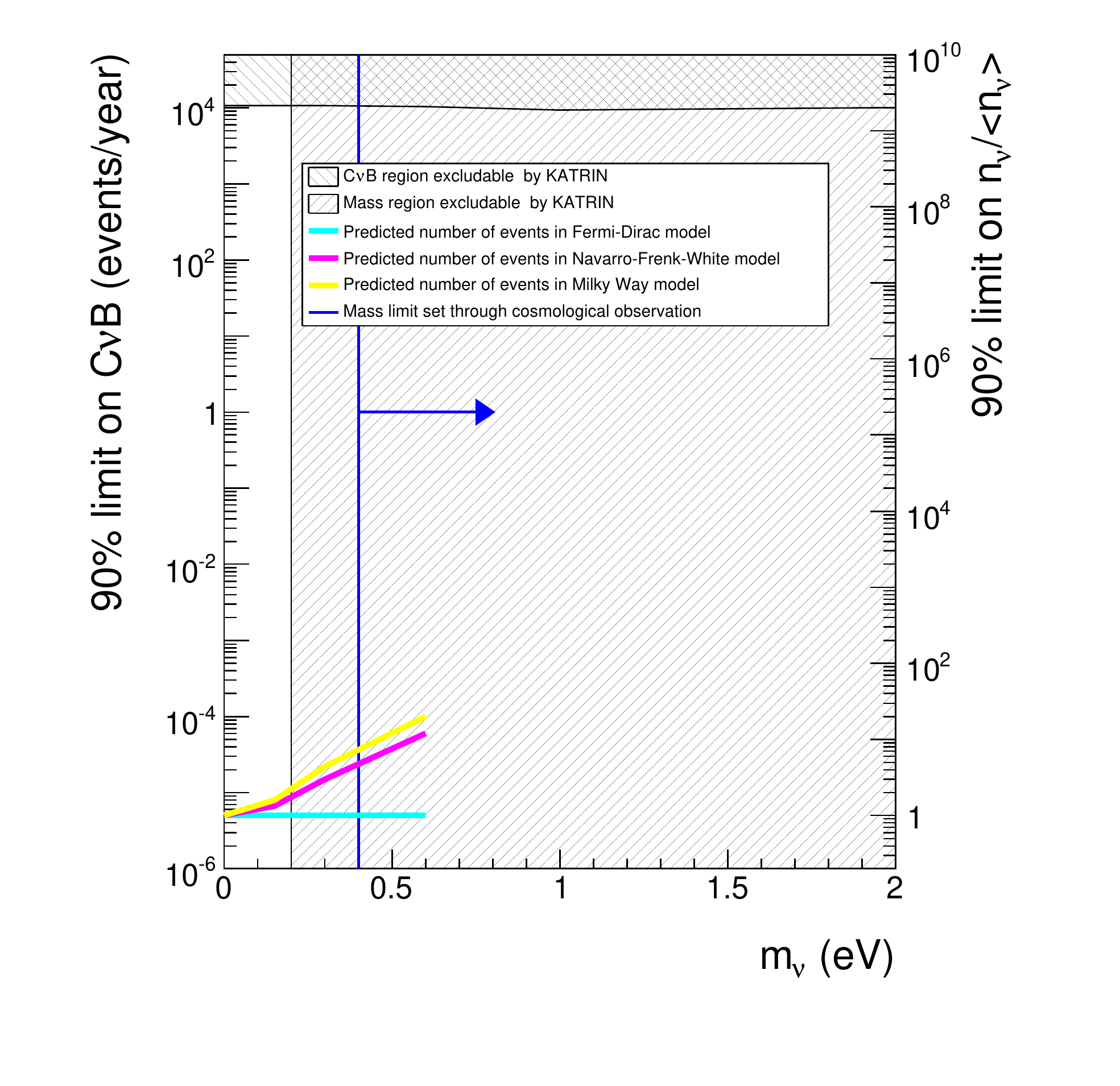} 
\caption{The 90\% confidence level sensitivity limit for relic neutrino over-density as a function of neutrino mass as expected from the 3 year data run at the KATRIN neutrino mass experiment. Solid curves show expectation from cosmological prediction assuming Fermi-Dirac (light blue), Navarro-Frenk-White (violet), and Milky Way (yellow) mass distribution.  Arrow shows neutrino mass limits already obtained from cosmological observations($\sum m_\nu \le$ 1.2 eV)~\cite{Fogli:2008jx}.}
\label{fig:limits}
\end{figure}

To calculate the sensitivity, the data sets were fit over six parameters: the neutrino mass ($m_\nu$), the \cnb~rate ($N_{C\nu B}$), the endpoint of the beta decay spectrum ($E_0$), the beta decay rate ($N_{T_2}$), the width of the \cnb~peak ($\sigma$), and the background rate ($N_b)$. We assume the same projected KATRIN run measurement plan as reported in~\cite{angrik2005kdr}. The results of the fit in the mass and capture rate are plotted for 2000 simulated experiments to create the confidence regions shown in Fig~\ref{fig:contours}. The figure shows only statistical errors; the systematic errors are estimated for four of the major errors for KATRIN: high voltage precisions, magnetic field precision, the effect of final states, and the error on the number of available tritium atoms. The errors are estimated by shifting the Monte Carlo data sets by one  standard deviation on the error and fitting at the central value of that parameter. The relative contributions to the errors for the mass and \cnb~rate are show in Table~\ref{tab:errors}.  The 90\% limit is shown for a variety of masses in Fig~\ref{fig:limits}. Shown on the right hand side of the plot is the limit on the local density of neutrinos at Earth. There is a slight decrease in sensitivity near 2 eV due in part because the run plan for KATRIN is discretized and optimized for a neutrino mass search. The discretization of 0.5 V in the region of interest means that the \cnb~search is restricted to a few points which do not change significantly with increasing mass up to about 1eV. However, above 1 eV, the finite endpoint of the default run plan means there are not sufficient bins to firmly establish the background level. Widening the energy scan from the original plan improves the limit significantly.

\begin{table}[htbp]
   \centering
   \begin{tabular}{|c|c|c|} 
	\hline
	Contributor&Error (events/year)&Percentage of Statistical\\
	\hline
	High Voltage& $\pm$ 5850& 70.1\%\\
	Magnetic Field& $\pm$ 2020& 24.2\%\\
	Final States& $\pm$ 1420& 17.0\%\\
	Normalization& $\pm$ 2080& 24.9\% \\
	\hline
	Statistical& $\pm$ 8340& -- \\
	\hline
	Total& $\pm$ 10680& 128\% \\
	\hline
   \end{tabular}
   \caption{Error contributions to the \cnb~for four major KATRIN systematics at m$_{\nu}=0$~eV. Errors are extended to other masses as a percentage of statistical errors. Note that the error on the final states is limited by Monte Carlo statistics.}
   \label{tab:errors}
\end{table}

From a limit on the local density at earth, certain theoretical possibilities can be investigated.  It has been proposed that the knee of the cosmic ray spectrum could be caused by a relic neutrino GZK effect~\cite{hwang2005detection}, which requires a neutrino density of greater than 10$^{13}$. Since the required overdensity to result in the limit KATRIN can set is 2.0$\times 10^{9}$, then, assuming KATRIN sees no signal at the expected sensitivity, KATRIN will exclude this model for neutrinos near earth. It has also been proposed that neutrinos could couple to one another via a light scalar boson and form bound clouds with significant effect on small scale structure formation in the universe~\cite{Stephenson:1996qj}. While this work shows that KATRIN is able to set a better limit on overdensity than the experiments considered therein by a factor of 10$^6$ (resulting in an improvement on the limit on the fermi momentum by a factor of 100), the ultimate limit on the coupling strength is also determined by the neutrino mass. In the range of masses accessible at KATRIN, the limit on the coupling strength could either entirely rule out this model or broaden the parameter space significantly. While no firm conclusions can be drawn with this work, it is definitely a topic for future analysis.

The direct observation of the cosmic neutrino background remains one of the strongest test of our cosmological framework.  The KATRIN neutrino mass experiment, via neutrino capture on tritium, can place a limit on the relic neutrino density of   $2.0 \times 10^{9}$ over standard model predictions at 90\% C.L., ruling out certain speculative cosmological models.

\section{Acknowledgements}

The authors would like to thank Janet Conrad, Peter Fisher, Andre deGouva, Klaus Eitel, and Hamish Robertson, as well as the members of the KATRIN collaboration, for fruitful discussions of the topic presented and the experimental challenges involved.  Joseph Formaggio and Asher Kaboth are supported by the U.S. Department of Energy under Grant No. DE-FG02-06ER-41420, while Benjamin Monreal is supported by the U.S. Department of Energy under Grant No. DE-SC0004036.

\bibliography{RelicWriteup_v4}

\end{document}